# Rule-based Autoregressive Moving Average Models for Forecasting Load on Special Days: A Case Study for France

Siddharth Arora[†] and James W. Taylor[*]

Saïd Business School,
University of Oxford, Park End Street, Oxford, OX1 1HP, U.K.

[†] Siddharth Arora
Tel: +44 (0)1865 288800
Email: Siddharth.Arora@sbs.ox.ac.uk

[*]James W. Taylor
Tel: +44 (0)1865 288800
Email: James.Taylor@sbs.ox.ac.uk

# Rule-based Autoregressive Moving Average Models for Forecasting Load on Special Days: A Case Study for France

**Abstract**

This paper presents a case study on short-term load forecasting for France, with emphasis on special days, such as public holidays. We investigate the generalisability to French data of a recently proposed approach, which generates forecasts for normal and special days in a coherent and unified framework, by incorporating subjective judgment in univariate statistical models using a rule-based methodology. The intraday, intraweek, and intrayear seasonality in load are accommodated using a rule-based triple seasonal adaptation of a seasonal autoregressive moving average (SARMA) model. We find that, for application to French load, the method requires an important adaption. We also adapt a recently proposed SARMA model that accommodates special day effects on an hourly basis using indicator variables. Using a rule formulated specifically for the French load, we compare the SARMA models with a range of different benchmark methods based on an evaluation of their point and density forecast accuracy. As sophisticated benchmarks, we employ the rule-based triple seasonal adaptations of Holt-Winters-Taylor (HWT) exponential smoothing and artificial neural networks (ANNs). We use nine years of half-hourly French load data, and consider lead times ranging from one half-hour up to a day ahead. The rule-based SARMA approach generated the most accurate forecasts.

# 1. Introduction

Accurate short-term forecasts of electricity demand (*load*) are crucial for making informed decisions regarding unit commitment, energy transfer scheduling, and load-frequency control of the power system. An electric utility needs to make these operational decisions on a daily basis, often in real-time, in order to operate in a safe and efficient manner, optimize operational costs, and improve the reliability of distributional networks. Moreover, inaccurate forecasts can have substantial financial implications for energy markets (Weron, 2006).

Given the significance of short-term load forecasts for electric utilities and energy markets, a plethora of different modelling approaches have been proposed for forecasting load for normal days (Bunn, 2000). Modelling load for special days, such as public holidays, however, has usually been overlooked in the research literature (see, for example, Hippert *et al.*, 2005; Taylor, 2010). Special days exhibit *load profiles* (shape of the intraday load curve) that differ noticeably from the repeating seasonal pattern that one might expect. We refer to load observed on normal days as *normal load*, whereas load observed on special days is referred to as *anomalous load*.

The lack of attention to modelling anomalous load can be attributed to the following: 1) Anomalous load deviates significantly from normal load, and is therefore not straightforward to model, as the special days need to be treated as being different from normal days. 2) The relatively infrequent occurrence of special days results in a lack of anomalous observations for adequately training the model. 3) Different special days exhibit different load profiles, which require each special day to be modelled as having a unique profile. The aforementioned reasons make statistical modelling of anomalous load very challenging, and this has tended to lead to the forecasting of anomalous load being left to the judgment of the central controller of the electricity grid (Hyde and Hodenett, 1993, 1997). The aim of this study is to investigate models that can potentially be deployed in a real-time automated online system, which can assist the central controller in making informed decisions under normal and anomalous load conditions.

The problem with ignoring anomalous load is that it not only guarantees that the resulting model cannot be used for special days, but it also results in large forecast errors on normal days that lie in the vicinity of special days. If special day effects are not modelled, there is seemingly a need either to replace or smooth observations for these days (see, for example, Smith, 2000; Hippert *et al.*, 2005; Taylor, 2010; Arora and Taylor, 2016). We use the actual load time series for modelling, whereby the anomalous observations are neither replaced, nor smoothed out.

Multivariate weather-based models have been employed previously for modelling load (Cottet and Smith, 2003; Dordonnat *et al.*, 2008; Cho *et al*., 2013). Multivariate models utilize weather variables like temperature, wind speed, cloud cover, and humidity, along with the historical load observations. Univariate models on the other hand, include only the historical load observations. It has been argued that the weather variables tend to vary smoothly over short time scales, and this variation can be captured in the load data itself (Bunn, 1982). Moreover, for short lead times, univariate models have been shown to be competitive with weather-based models (Taylor, 2008). In this paper, we employ univariate methods for short-term load forecasting.

Rule-based forecasting has been proposed as a practical way to incorporate subjective judgment, based on domain knowledge and expertise, into a statistical model (see, for example, Armstrong, 2001, 2006). It has been argued that rule-based forecasting can outperform conventional extrapolation methods, especially in cases where prior domain knowledge is available and the time series exhibits a consistent structure (see, for example, Collopy and Armstrong, 1992; Adyaa *et al.*, 2000). Given that the task of forecasting anomalous load has previously relied mainly on subjective judgment, and the fact that load exhibits a consistent prominent seasonal structure, we adopt a rule-based methodology in this study. We incorporate domain knowledge into the statistical models via a rule. The rationale of the proposed rule lies in identifying an historical special day, whose anomalous load observations would be most useful in improving the accuracy of the model in forecasting load for the future special day.

In a recent study, Arora and Taylor (2013) propose rule-based approaches for modelling anomalous load for Great Britain. Of the methods considered, the most successful was based on SARMA modelling. In this paper, we focus on the use of this method for modelling French load data. In comparison with load for Great Britain, modelling anomalous load for France is more challenging, due to the relatively large number of different types of special days observed in France. As a consequence of this, for the French case, the approach of Arora and Taylor (2013) requires an adaptation that, although methodologically modest, is important empirically.

The contributions of this study lie in: 1) Presenting a detailed case study for France, which focusses on short-term forecasting of anomalous load using a range of different modelling approaches. 2) Formulating a rule specifically for the French load data that allows for incorporation of domain knowledge into the statistical framework during the modelling process. Crucially, the formulated rule treats each special day as having a unique profile, which may change over the years. The rule categorizes special days into seven different categories based on an inspection of the anomalous load profiles. 3) Adapting a double seasonal SARMA method recently proposed in this journal for anomalous load forecasting of Korean data (see Kim, 2013). In our adaptation of Kim's method, we treat each special day as having a unique profile, we incorporate an additional dummy variable in the model to allow for greater flexibility in accommodating special day effects, and we model triple seasonality.

Moreover, we propose a range of different benchmarks for assessing load forecasts on special days. 4) In addition to generating point forecasts, we evaluate probability density forecasts across normal and special days. To the best of our knowledge, there are no existing studies on density forecasting for anomalous load.

In the next section, we present the French load data. In Section 3, we review the literature on modelling anomalous load. In Section 4, we present the rule-based SARMA method along with the subjective formulation of a rule. Section 5 presents an adaption of the SARMA model proposed by Kim (2013). In Section 6, we present the benchmark methods. Empirical comparison is provided in Section 7. In Section 8, we summarise and conclude the paper.

## 2.    Anomalous Load Characteristics

We employ nine years of half-hourly load for France, stretching from 1 January 2001 to 31 December 2009, inclusive. This leads to a total of 157,776 load observations. We use the first eight years of the dataset as the estimation sample (consisting of 140,256 observations), and employ the final year as the evaluation sample (consisting of 17,520 observations). We generate forecasts by rolling the forecast origin through each half-hour in the post-sample period. The data has been obtained from Électricité de France (EDF), and there are no missing observations in the time series. The complete series is presented in Figure 1. It can be seen from this figure that load exhibits a recurring within-year pattern (due to seasonal effects), termed the intrayear seasonality. Load in winter is higher than in summer, which is due to the increased use of electrical equipment for winter heating in France. Also, the data shows an upward trend.

The average intraday cycle for different days of the week (calculated using only the estimation sample) is presented in Figure 2. It can be seen from the figure that load on weekends is considerably lower compared to weekdays, and load is lowest on Sundays. Load on Monday mornings and Friday evenings is lower than other weekdays, whereas the average load profiles for Tuesday, Wednesday and Thursday are very similar.

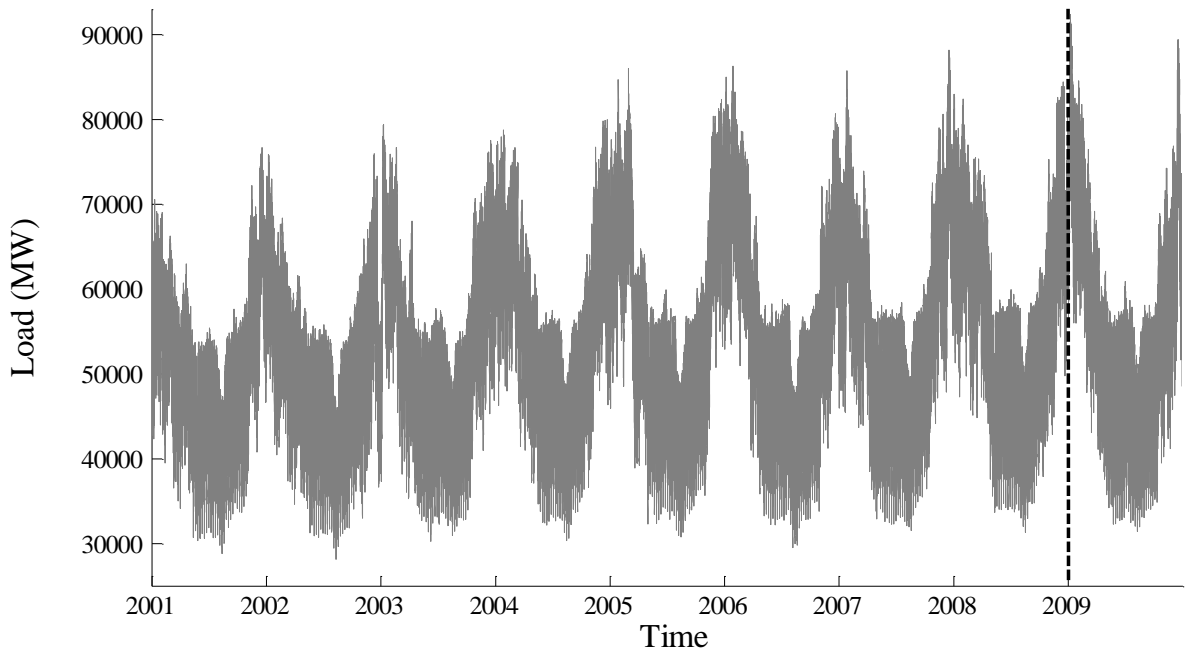

**Figure 1**— Half-hourly load for France from 1 January 2001 to 31 December 2009. The vertical dashed line divides the time series into estimation and evaluation samples.

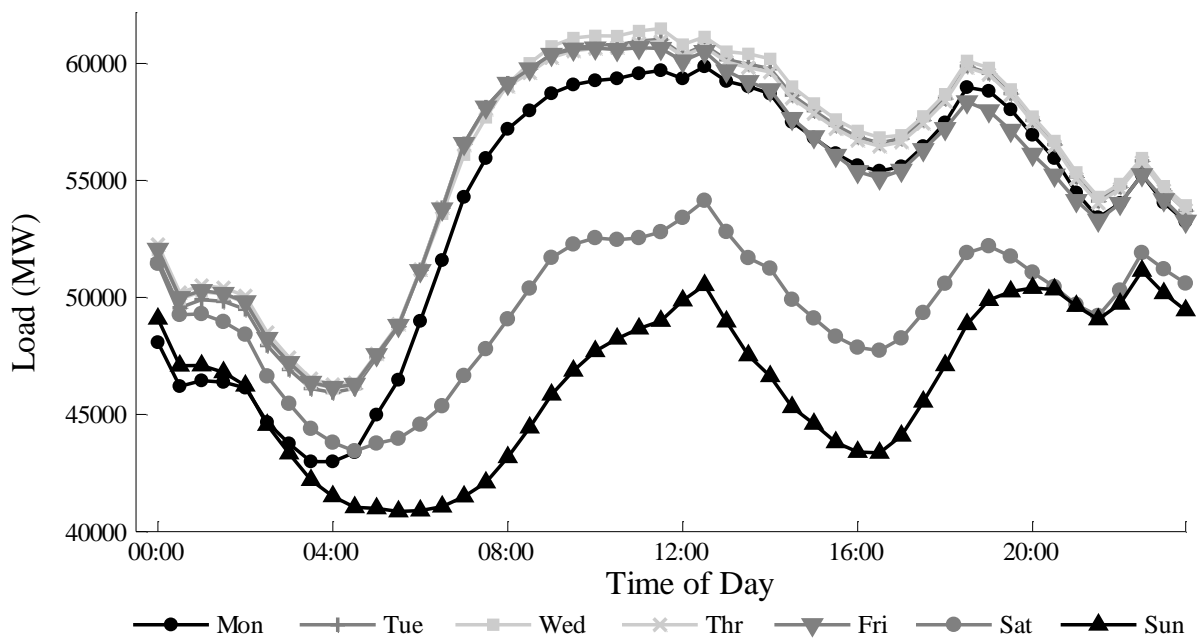

**Figure 2**— Average intraday profile for each day of the week.

We identify a total of twenty four special days in the one year post-sample period. Inspection of the data reveals that load for a given special day is considerably lower than a normal day, for the same day of the week, around the same date. Figure 3 compares anomalous and normal load. In this figure, we plot load for Bastille Day in 2008 (14 July 2008), which is a national holiday in France and which fell on a Monday that year. In the figure, we also plot load for a normal working Monday from the preceding and following weeks. It is evident from Figure 3 that, not only is anomalous load substantially lower than normal load, but the shape of the load profiles for normal and special days are indeed very different. Inspection of the data reveals that this characteristic of anomalous load holds true across all special days.

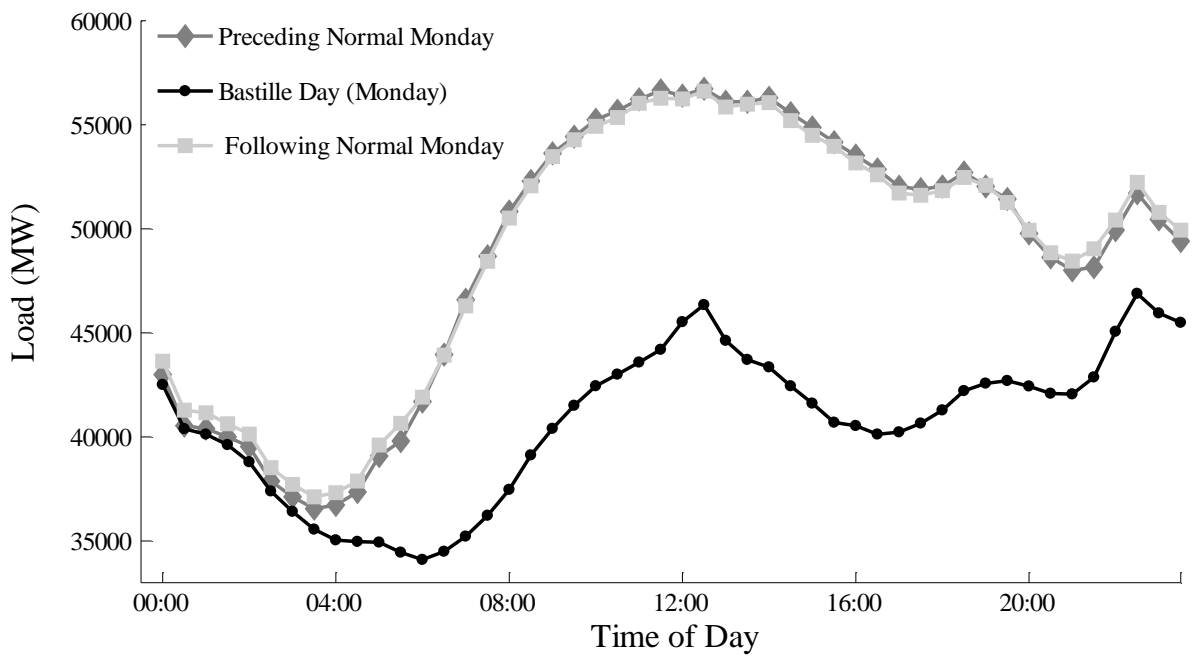

**Figure 3**— Load profile for a Bastille Day, which fell on a Monday (14 July 2008), a normal Monday (7 July 2008) from the preceding week, and a normal Monday (21 July 2008) from the following week.

In Figure 4, we plot load profiles for six different special days observed in the year 2008. As expected, load for special days occurring during winter months (New Year's Day, Christmas Day and Remembrance Day) is considerably higher than load for special days occurring during summer months (Labor Day, Bastille Day and the Assumption). However, it is also interesting to note that the profile shape differs for different special days. For example, demand on Labor Day and Bastille Day differ substantially at the start of the day, but the difference is much less late in the day.

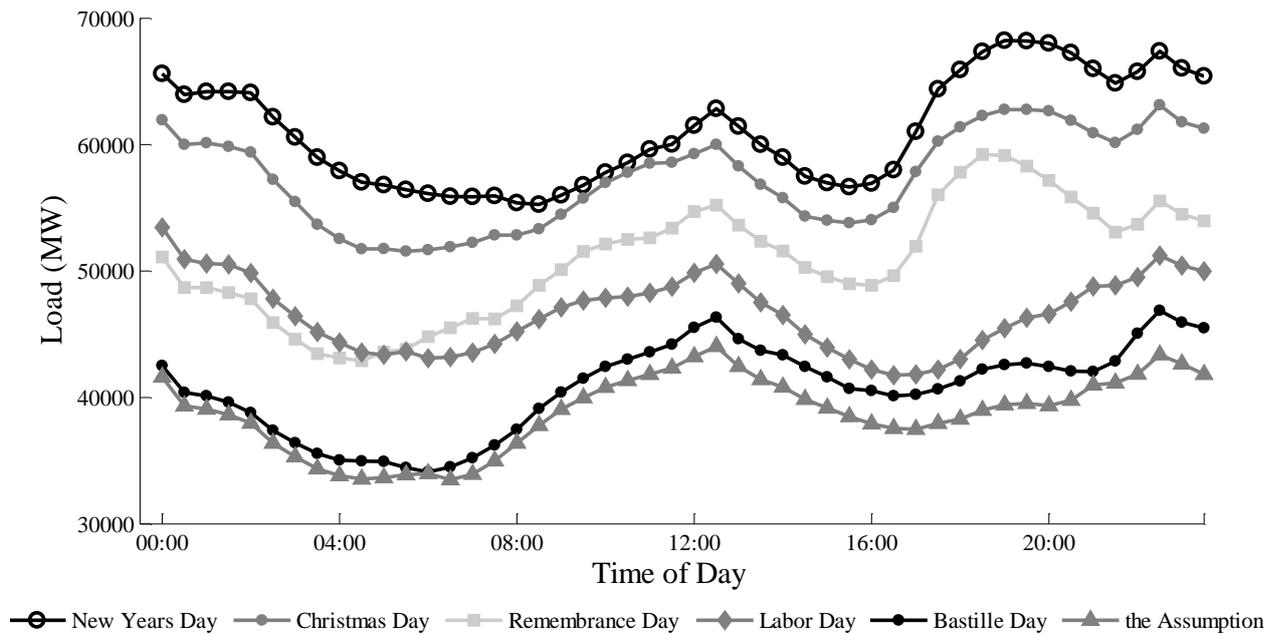

**Figure 4**— Load profile for New Year's Day (1 January), Christmas Day (25 December), Remembrance Day (11 November), Labor Day (1 May), Bastille Day (14 July), and the Assumption (15 August), observed in the year 2008.

In Figure 5, for the eight years of our estimation sample (2001-2008), we plot load for New Year's Day (1 January), and a normal day occurring on the same day of the week a fortnight later (15 January). It can be seen from Figure 5, that the profile for a given special day is different across different years, and the same is true for a normal day. Moreover, we note that the upward trend that is apparent in Figure 1 is not so clear in Figure 5, and indeed the relative levels of load in the eight years is not the same for the special days and normal days.

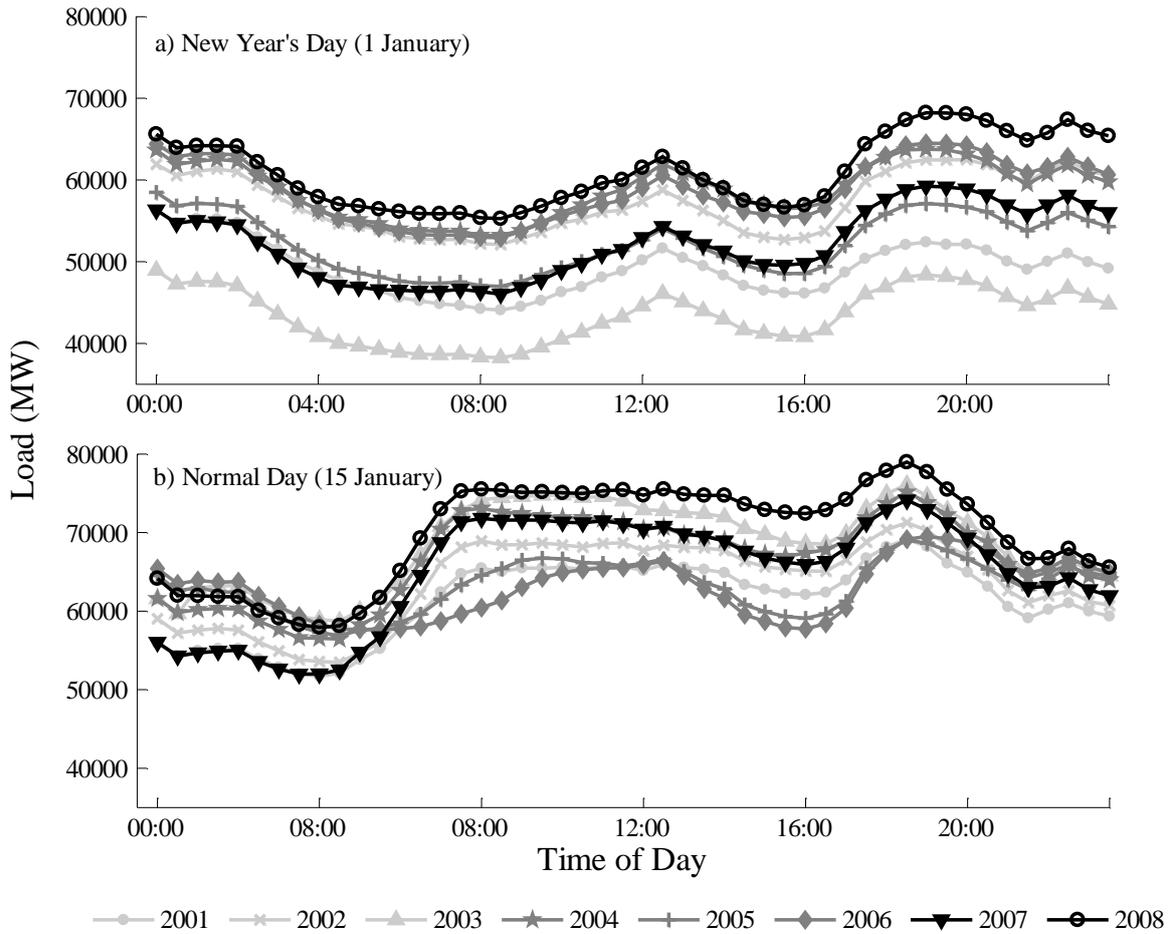

**Figure 5**— Load profile for New Year's Day (1 January), and a normal day (15 January) observed across all years within the estimation sample, 2001 to 2008.

Several researchers have incorporated proximity day effects while modelling anomalous load (see, for example, Ramanathan *et al.*, 1997, Pardo *et al.*, 2002; Dordonnat *et al.*, 2008; Kim, 2013; Arora and Taylor, 2013). The day which either precedes or follows a special day is defined as a proximity day. Due to special day effects, load on proximity days tends to be lower than normal load for the same day of the week (around the same date), but higher than the corresponding special day.

In Figure 6, we plot load for a special day (1 May, Labor Day, Thursday), a proximity day (2 May, Friday) that follows the special day, and a corresponding normal day from the previous week (24 April, Thursday), all observed in the year 2008. It is evident from the figure that load on the proximity day is noticeably lower than normal load, but considerably higher than anomalous load.

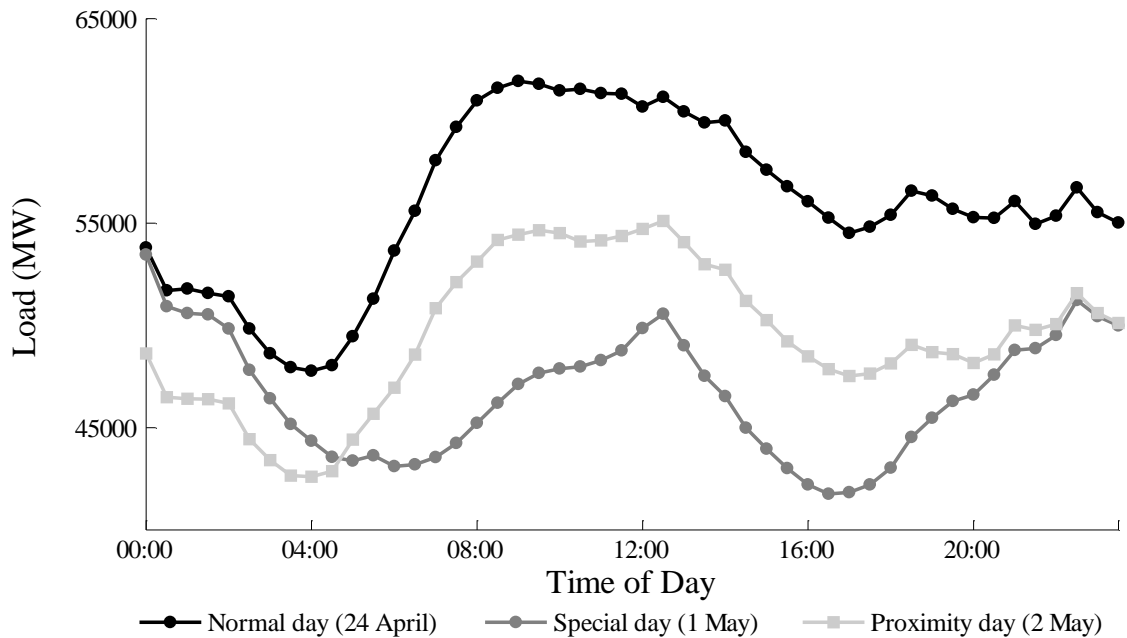

**Figure 6**— Load profile for a normal day (24 April, Thursday), special day (1 May, Labor Day, Thursday), and a proximity day (2 May, Friday), observed in the year 2008.

We identify proximity days based on visual inspection of the data, and treat them as either *bridging* or *non-bridging days*, as explained in detail in Section 4. A proximity day is defined to be a bridging day if it is a Monday before a special day, or a Friday after a special day. Otherwise, we define proximity day as a non-bridging day. The load on bridging days tends to be lower than non-bridging days, and hence, needs to be treated separately.

## 3.     Review of Methods for Anomalous Load Forecasting

Previous approaches for short-term forecasting of normal and anomalous load have mostly employed regression-based methods with dummy variables for special days. Ramanathan *et al.* (1997) build a separate regression model for each hour of the day, and include special day and weather effects using dummy variables. Pardo *et al.* (2002) use dummy variables for capturing the day of week, month of year, and holiday effects, and employ additional dummy variables for the day following a special day, as these days often also exhibit abnormal load patterns. Cottet and Smith (2003) model load using a multi-equation Bayesian model, whereby they employ 48 coefficients for each dummy variable to capture the intraday seasonality in their half-hourly load data. Cancelo *et al.* (2008) build a separate model for each hour of the day using Spanish load. They first issue a forecast for load assuming a normal day, and make adjustments accordingly for special days using different dummy variables employed for different classes of special days. Soares and Medeiros (2008) build a two-stage model for each hour of the day, such that anomalous load is modelled in the first stage using dummy variables. Any unexplained component in load is then modelled in the second stage using either an AR model or an ANN. Dordonnat *et al.* (2008) build a regression model for each hour of the day, and accommodate special day effects using dummy variables. In addition, they also use dummy variables for bridging days. Gould *et al*. (2008) propose a state space approach for forecasting time series with multiple seasonal patterns, and accommodate special day effects by treating them as if they are Saturdays or Sundays. De Livera *et al.* (2011) propose an innovations state space modelling framework, and handle special day effects for national and religious holidays in load via dummy variables. Kim (2013) employs a double seasonal ARMA model to accommodate special day effects on an hourly basis using indicator variables for Korean load. In Section 5, we adapt the method used by Kim (2013) for French load.

For the approach to anomalous load that involves the use of regression models and dummy variables, to avoid over-parameterisation, different special days have often been classified as belonging to the same special day type (see, for example, Kim *et al*., 2000; Cottet and Smith, 2003; Cancelo *et al.*, 2008; Soares and Medeiros, 2008; Dordonnat *et al*., 2008; Kim, 2013). The classification of special days relies on the assumption that the load profile for different special days can be treated as being similar, and would remain similar over the years. Using French load, we observe from the data that each special day exhibits a unique profile, as shown in Figure 4. Hence, instead of classifying different special days as being the same, we model each special day as having a unique profile, which may change over different years.

Most existing methods for anomalous load forecasting rely on classifying different special days as being the same, while some approaches employ different models for normal and special days (Kim *et al*., 2000). By contrast, Arora and Taylor treat each special day as having a unique profile, and adopt a unified modelling framework for normal and special days. In this paper, we adapt and apply the approach of Arora and Taylor to French load.

Apart from regression-based methods, some authors have proposed rule-based approaches for anomalous load forecasting, while others have used ANNs. Rahman and Bhatnagar (1988) propose a rule-based approach, whereby they formulate rules based on the logical and syntactical relationships between weather and load. Hyde and Hodnett (1997) formulate rules for Irish load data, whereby the rationale of their approach is to find the deviation of load for different special days from normal load for a given year, and use this deviation as a correction term for the corresponding special day falling next year. Kim *et al*. (2000) classify special days into five different types, and employ an ANN (for each special day type) used in conjunction with fuzzy rules inferred from their Korean load data. Recently, Barrow and Kourentzes (2016) proposed ANNs for accommodating the effect of special days while modelling call centre arrivals. Using load time series for Great Britain, Arora and Taylor (2013) demonstrate how a set of univariate methods can be adapted to model load for special days, when used in conjunction with a rule-based approach.

The existing rule-based methods are tailored only to the data at hand (Rahman and Bhatnagar, 1988; Hyde and Hodnett, 1993, 1997). This makes the task of adapting existing rule-based methods to different datasets very challenging, and would require, for these rule-based methods, creating a completely new set of rules for the French data. For this reason, we do not use existing rule-based methods as benchmark methods in this study.

## 4. Rule-based Modelling Framework

This section presents a rule-based SARMA method, along with the principles of formulating a rule based on a categorization of special days. Our methodological contribution to this section is to provide an important adaption of the rule for application to the French case.

### 4.1. Rule-based SARMA

In this paper, for the French load data, we consider the rule-based adaptation of SARMA proposed by Arora and Taylor (2013) for modelling anomalous load. The model has the following formulation:

$$Y_p(L)\Phi_{P_1}(L^{m_1})X_{P_2}(L^{m_2})\left(I_{N_t}\Psi(L^{m_3(t)})+(1-I_{N_t})\theta(L^{m_3(t)})\right)(y_t - c) = \\ \Omega_q(L)\Theta_{Q_1}(L^{m_1})\Gamma_{Q_2}(L^{m_2})\left(I_{N_t}\Lambda(L^{m_3(t)})+(1-I_{N_t})K(L^{m_3(t)})\right)\left(I_{N_t}\varepsilon_t^{(N)}+(1-I_{N_t})\varepsilon_t^{(S)}\right) \quad (1)$$

where $y_t$ denotes the load observed at period $t$, $c$ is a constant parameter; $L$ denotes a lag operator; and $Y_p$, $\Phi_{P_1}$ and $X_{P_2}$ are AR polynomial functions of order $p$, $P_1$ and $P_2$, while $\Omega_q$, $\Theta_{Q_1}$ and $\Gamma_{Q_2}$ are MA polynomial functions of order $q$, $Q_1$ and $Q_2$, respectively. We consider polynomial function orders equal to or less than three. The model errors for normal and special days are denoted by $\varepsilon_t^{(N)} \sim NID(0, \sigma_N^2)$ and $\varepsilon_t^{(S)} \sim NID(0, \sigma_S^2)$, respectively, having corresponding variances $\sigma_N^2$ and $\sigma_S^2$, while $NID$ refers to a normal and independently distributed process. For any period $t$ occurring on a normal day, the binary indicator $I_{N_t}$ equals one, and zero otherwise.

The length of the intraday, intraweek, and intrayear seasonal cycle are denoted by $m_1$, $m_2$ and $m_3(t)$. Since the data is recorded every half-hour, we have $m_1 = 48$ and $m_2 = 336$. For normal days, we have $m_3(t) = 52 \times m_2$ (for a given period $t$), except for a few weeks around the clock-change, where $m_3(t) = 53 \times m_2$. (Clock-change involves the clocks being put forward by an hour on the last Sunday in March, and put back one hour on the last Sunday in October.) For special days, it is very important to note that $m_3(t)$ is selected using a rule-based approach, which we describe in detail later in this section. We refer to this model as rule-based triple seasonal autoregressive moving average (RB-SARMA).

In expression (1), the functions $\Psi$ and $\Lambda$ accommodate the intrayear seasonal effects for normal days; while $\theta$ and $K$ accommodate intrayear seasonality for special days. The function $\Psi$ is written as:

$$\Psi(L^{m_3(t)}) = 1 + \eta_1 L^{m_3(t)} + \eta_2 L^{m_3(t)+m_3(t-m_3(t))} + \eta_3 L^{m_3(t)+m_3(t-m_3(t))+m_3(t-m_3(t-m_3(t)))} \quad (2)$$

where $\eta_1, \eta_2$ and $\eta_3$ are constant parameters. The functions $\theta$, $\Lambda$ and $K$ are of the same form as $\Psi$, with the difference that each function involves a different set of parameters. With these functions, the formulation in expression (1) involves switching between different AR and MA polynomial functions, with different annual lag terms, depending on whether $y_t$ belongs to a normal day, or a special day. We adopt the Box and Jenkins (1970) methodology to select the orders for all the polynomial functions in expression (1).

The model parameters for RB-SARMA are estimated by maximum likelihood, employing only the in-sample data. The likelihood function assumes a Gaussian error, for which the variance is different on special days to that of normal days. Specifically, we use the following log-likelihood (*LL*) function:

$$LL = -\frac{n_{ND}}{2}\log(2\pi\sigma_N^2) - \frac{n_{SD}}{2}\log(2\pi\sigma_S^2) - \sum_{t=365 \times m_1 + 1}^{N} \left( \frac{I_{N_t}}{2\sigma_N^2}(\varepsilon_t^{(N)})^2 + \frac{(1-I_{N_t})}{2\sigma_S^2}(\varepsilon_t^{(S)})^2 \right) \quad (3)$$

where $N$ is the length of the estimation sample, $n_{ND}$ and $n_{SD}$ are the number of load observations that belong to normal and special days, respectively, excluding the observations from the first year. To estimate model parameters, we used an optimization scheme based on a simplex search method of Lagarias *et al.* (1998). Once estimated, the parameters were held fixed, and the estimated model was employed for generating forecasts for the out-of-sample data.

*4.2. Rule Formulation: Principles*

The RB-SARMA method described in Section 4.1 requires the value of the intrayear cycle length, $m_3(t)$, to be specified for each period $t$ in the estimation sample, and each period for which a forecast is required. For normal days, as we stated in Section 4.1, $m_3(t)$ is set to be either $52 \times m_2$ or $53 \times m_2$ (depending on the proximity of $t$ to the clock-change). For special days, each value for $m_3(t)$ is essentially chosen subjectively. This subjectivity is supported by a rule-based framework, which is the focus of this section. The important point to appreciate is that the sole purpose of the rule is to determine, for each special day, the value of $m_3(t)$. In our description of the rule, we refer to the day on which period $t$ falls as the *current special day*, and the day on which $t$-$m_3(t)$ falls as the *corresponding past special day*.

The rule considered in this study is an adaptation of a rule proposed by Arora and Taylor (2013). In that study, four different rules were proposed for a British load time series. Of the four rules, Rule 3 performed the best for special days. In this paper, we focus on this one rule, and adapt it for the relatively large number of different types of special days observed in France. We ensure that the rule treats each special day as having a unique profile. Specifically, for each special day, $m_3(t)$ is chosen in accordance with the following principles:

(i) $m_3(t)$ is chosen to be a multiple of $m_1$, which is equal to 48 for our French data. This seems intuitively reasonable, as it ensures that $m_3(t)$ relates each period in the current special day to the same period of the day in the corresponding past special day.

(ii) $m_3(t)$ is chosen to be the same for all periods $t$ on the current special day. This means that, for each current special day, the rule prescribes a single corresponding past special day.

(iii) Figure 2 shows that the average intraday load profile for weekends is substantially lower compared to weekdays. In view of this, for each special day, $m_3(t)$ is chosen so that the current day and corresponding past day are either both weekdays, or both fall on weekends.

(iv) When choosing between two past special days from different years, to use as the corresponding past special day, select the one from the most recent of the two years.

(v) The corresponding past special day should be chosen so that it occurs at a similar time of year to the current special day. For some special days, the date of current and corresponding past special days should be the same. For others, it is more important that the day of the week is the same. The difference in how the corresponding past special day should be selected leads us to categorise the special days into one of seven different special day types. Specifically, it is important to note that to model load for a given special day, we refer to the most recent past special day that belongs to the same *category* as the special day under consideration. In defining these categories, we incorporate all of the principles described in this section. We present the seven categories in Section 4.3.

*4.3. Rule Formulation: Special Day Categories*

Table 1 presents the special days of 2009, with each allocated to one of the seven categories of special day types. For each special day in 2009, the table also shows the corresponding past special day. In this section, we describe the categorization of special days.

We define the national public holidays as *basic special days*. Each seems to have a somewhat different load profile to any other special day, and so in defining $m_3(t)$, it seems sensible to relate each of these special days to the same special day occurring in the past. With patterns of load on weekdays being very different to patterns on weekends, if the current basic special day falls on a weekday, $m_3(t)$ is chosen so that the corresponding past day is chosen as the most recent occurrence of the same special day that fell on a weekday. Similarly, if the current basic special day falls on a weekend, $m_3(t)$ is set in order that the corresponding past day is chosen as the most recent occurrence of the same special day that fell on a weekend. This leads us to the following two categories of special days:

**Category A**: Basic special days that occur on a weekday.

**Category B**: Basic special days that occur on a weekend.

For 2009, Table 1 presents the 10 special days in Category A and the 3 special days in Category B, and their corresponding past special days.

As we explained in Section 2, a proximity day is a special day that either precedes or follows a basic special day. Load on all proximity days is sufficiently abnormal that it is necessary to treat them as special days. The pattern of load on a proximity day that follows a special day is typically different to the load profiles for proximity days that precede a special day. This is reflected in the categorization. Furthermore, the categorization also accounts for proximity days that are bridging days. As described in Section 2, a bridging day is a proximity day that occurs between a special day and a weekend. Load on a bridging day tends to be lower than on non-bridging proximity days. Considerations regarding proximity days lead us to the following five categories of special days:

**Category C**: Bridging proximity days that precede a special day.

**Category D**: Bridging proximity days that follow a special day.

**Category E**: Non-bridging proximity days that precede a special day, and occur on a weekday.

**Category F**: Non-bridging proximity days that follow a special day, and occur on a weekday.

**Category G**: Non-bridging proximity days that follow a special day, and occur on a weekend.

TABLE 1: CATEGORISATION OF FRENCH SPECIAL DAYS IN 2009, AND THE CORRESPONDING PAST SPECIAL DAY SELECTED TO DEFINE $m_3(t)$.

| **Current Special Day :** | | **Corresponding Past Special Day Referred to via $m_3(t)$** | |
|---|---|---|---|
| *Category A: Basic special days that occur on a weekday* | | | |
| New Year's Day : *Thu* | 01/01/2009 | New Year's Day : *Tue* | 01/01/2008 |
| Easter Monday : *Mon* | 13/04/2009 | Easter Monday : *Mon* | 24/03/2008 |
| Labor Day : *Fri* | 01/05/2009 | Labor Day : *Thu* | 01/05/2008 |
| WWII Victory Day : *Fri* | 08/05/2009 | WWII Victory Day : *Thu* | 08/05/2008 |
| Ascension Day : *Thu* | 21/05/2009 | Ascension Day : *Thu* | 17/05/2007 |
| Whit Monday : *Mon* | 01/06/2009 | Whit Monday : *Mon* | 12/05/2008 |
| Bastille Day : *Tue* | 14/07/2009 | Bastille Day : *Mon* | 14/07/2008 |
| Remembrance Day : *Wed* | 11/11/2009 | Remembrance Day : *Tue* | 11/11/2008 |
| Christmas Day : *Fri* | 25/12/2009 | Christmas Day : *Thu* | 25/12/2008 |
| New Year's Eve : *Thu* | 31/12/2009 | New Year's Eve : *Wed* | 31/12/2008 |
| *Category B: Basic special days that occur on a weekend* | | | |
| * The Assumption : *Sat* | 15/08/2009 | The Assumption : *Fri* | 15/08/2008 |
| All Saints Day : *Sun* | 01/11/2009 | All Saints Day : *Sat* | 01/11/2008 |
| Boxing Day : *Sat* | 26/12/2009 | Boxing Day : *Sun* | 26/12/2004 |
| *Category C: Bridging proximity days that precede a special day* | | | |
| ** Day before Bastille Day : *Mon* | 13/07/2009 | Day after Bastille Day : *Fri* | 15/07/2005 |
| *Category D: Bridging proximity days that follow a special day* | | | |
| Day after New Year's : *Fri* | 02/01/2009 | Day after New Year's : *Fri* | 02/01/2004 |
| Day after Ascension : *Fri* | 22/05/2009 | Day after Ascension : *Fri* | 18/05/2007 |
| *Category E: Non-bridging proximity days that precede a special day, and occur on a weekday* | | | |
| Christmas Week : *Mon* | 21/12/2009 | Christmas Week : *Mon* | 22/12/2008 |
| Christmas Week : *Tue* | 22/12/2009 | Christmas Week : *Mon* | 22/12/2008 |
| Christmas Week : *Wed* | 23/12/2009 | Christmas Week : *Tue* | 23/12/2008 |
| Christmas Week : *Thu* | 24/12/2009 | Christmas Week : *Wed* | 24/12/2008 |
| *Category F: Non-bridging proximity days that follow a special day, and occur on a weekday* | | | |
| Christmas Week : *Mon* | 28/12/2009 | Christmas Week : *Mon* | 29/12/2008 |
| Christmas Week : *Tue* | 29/12/2009 | Christmas Week : *Mon* | 29/12/2008 |
| Christmas Week : *Wed* | 30/12/2009 | Christmas Week : *Tue* | 30/12/2008 |
| *Category G: Non-bridging proximity days that follow a special day, and occur on a weekend* | | | |
| Christmas Week : *Sun* | 27/12/2009 | Christmas Week : *Sat* | 30/12/2006 |

* The Assumption in 2009 occurred on a weekend (Saturday), however, there were no instances of this basic special day falling on a weekend in the estimation period, hence, for this case, we simply refer to the profile of the same special day from the previous year (2008).

** For the bridging proximity day that preceded Bastille Day in 2009 (Monday), we did not have an instance of a corresponding past bridging proximity day that preceded Bastille Day and fell on same day of the week (Monday), hence, for this case, we simply refer to the profile of a past bridging proximity day that follows Bastille Day.

Let us provide an illustrative example of the specification of $m_3(t)$ for a proximity day. Let us consider Friday 22 May 2009. This day is allocated to Category D, because it is a Friday following Ascension Day (Thursday 21 May 2009). The rule requires us to select, as corresponding past special day, the special day in Category D that occurred at the same time of year, most recently. This leads to the choice of Friday 18 May 2007, as the corresponding past special day, because Ascension Day occurred on 17 May in 2007. Hence, for all periods on Friday 22 May 2009, we set $m_3(t) = (365 + 366 + 4) \times 48$. In cases where the magnitude of $m_3(t)$ is larger than the total number of historical observations, the corresponding special day is set as the same special day from the previous year.

## 5. SARMA with Indicator Variables for Special Days

In a recent study in this journal, Kim (2013) used a SARMAX model (double seasonal ARMA with indicator variables) to forecast anomalous load. Specifically, Kim (2013) employed two types of indicator variables (denoted by $A_{h,t}$ and $B_{h,t}$) in the SARMAX modelling framework. The variable $A_{h,t}$ was an indicator variable used to indicate whether load observed on intraday period $h$ on a special day (for a given period $t$) deviates from normal load, whereas $B_{h,t}$ was employed to quantify the extent of this deviation.

Kim (2013) treated different special days as being the same, with one variable ($B_{h,t}$) to distinguish between different levels of deviation between normal and anomalous load. In this study, we implement an adaptation of the model, which, in our empirical analysis, improved forecast accuracy considerably. In our adaptation, we: 1) treat each special day as having a unique profile; 2) employ an additional indicator variable ($C_t$) to enable greater flexibility in accommodating the special day effects; and 3) model triple seasonality. We refer to the adapted model as SARMAX(ABC), and formulate it as:

$$y_t = c + \sum_{h=1}^{m_1} \alpha_h A_{h,t} + \sum_{h=1}^{m_1} \beta_h B_{h,t} + \sum_{h=1}^{m_1} \gamma_h C_{h,t} \qquad (4)$$
$$+ \frac{\Omega_q(L)\Theta_{Q_1}(L^{m_1})\Gamma_{Q_2}(L^{m_2})\Lambda(L^{m_3(t)})}{\Upsilon_p(L)\Phi_{P_1}(L^{m_1})X_{P_2}(L^{m_2})\Psi(L^{m_3(t)})}\left(I_{N_t}\varepsilon_t^{(N)}+(1-I_{N_t})\varepsilon_t^{(S)}\right)$$

where $h$ is counter for the $m_1$ periods of in each day; $\alpha_h$, $\beta_h$ and $\gamma_h$ are the model parameters; $A_{h,t}$, $B_{h,t}$ and $C_{h,t}$ are the indicator variables. For normal and anomalous periods, we use $m_3(t) = 52 \times m_2$ (except for a few weeks around clock-change, where $m_3(t) = 53 \times m_2$). The remaining terms for SARMAX(ABC) are as defined earlier for RB-SARMA.

To understand the indicator variables, $A_{h,t}$, $B_{h,t}$ and $C_{h,t}$, first note that each is equal to zero if period $t$ does not occur on intraday period $h$. If period $t$ does fall on intraday period $h$, the indicator variables take a value of 1 if that period is deemed to differ notably from what one might expect. The indicator variable values for all anomalous periods in 2009 are provided in Table 2.

To explain the indicator variable coding scheme more precisely, for load observed on a given special day ($y_t$), we refer to load observed on the same special day from the last year (denoted by $y_{t-m'_3(t)}$). We compute the percentage difference between $y_{t-m'_3(t)}$ and the mean of the previous 4 corresponding periods on normal days that fell on same day of the week ($\mu_t = \frac{1}{4}\sum_{i=1}^{4} y_{t-m'_3(t)-i\times m_1}$). Kim (2013) set $B_{h,t}=1$, if the percentage difference was between 10% and 20%, and $B_{h,t}=2$, if the percentage difference was greater than 20%, but we found this to be overly restrictive, and hence we included a third indicator variable $C_{h,t}$. We set $A_{h,t}=1$ if the percentage difference is at least 10%; we set $B_{h,t}=1$ if the percentage difference is between 10% and 20%; and we set $C_{h,t}=1$ if the percentage difference is more than 20%.

TABLE 2: CATEGORISATION OF FRENCH SPECIAL DAYS IN 2009, AND THE VALUE OF CORRESPONDING INDICATOR VARIABLES FOR DIFFERENT PERIODS OF THE DAY.

| Current Special Day: | | Intraday periods on which indicator variables are 1 | |
|---|---|---|---|
| | | $A_{h,t}=1; B_{h,t}=1$ | $A_{h,t}=1; C_{h,t}=1$ |
| *Category A: Basic special days that occur on a weekday* | | | |
| New Year's Day : *Thu* | 01/01/2009 | 05:30-06:30, 10:30-19:00 | - |
| Easter Monday : *Mon* | 13/04/2009 | - | - |
| Labor Day : *Fri* | 01/05/2009 | 02:30-05:00, 21:00-23:30 | 05:30-20:30 |
| WWII Victory Day : *Fri* | 08/05/2009 | 00:00-02:00, 21:30-23:30 | 02:30-21:00 |
| Ascension Day : *Thu* | 21/05/2009 | 02:30-05:00, 21:00-23:30 | 05:30-20:30 |
| Whit Monday : *Mon* | 01/06/2009 | 00:00-03:00, 21:30-23:30 | 04:00-21:00 |
| Bastille Day : *Tue* | 14/07/2009 | 05:00-06:00, 12:30, 19:30-23:00 | 06:30-12:00, 13:00-19:00 |
| Remembrance Day : *Wed* | 11/11/2009 | 05:30-06:30, 09:00-17:00 | 07:00-08:30 |
| Christmas Day : *Fri* | 25/12/2009 | 01:00-05:00, 20:00-23:30 | 05:30-19:30 |
| New Year's Eve : *Thu* | 31/12/2009 | - | - |
| *Category B: Basic special days that occur on a weekend* | | | |
| The Assumption : *Sat* | 15/08/2009 | 00:00-06:00, 19:30-23:30 | 06:30-19:00 |
| All Saints Day : *Sun* | 01/11/2009 | - | - |
| Boxing Day : *Sat* | 26/12/2009 | 00:00-10:30 | - |
| *Category C: Bridging proximity days that precede a special day* | | | |
| Day before Bastille Day : *Mon* | 13/07/2009 | - | - |
| *Category D: Bridging proximity days that follow a special day* | | | |
| Day after New Year's : *Fri* | 02/01/2009 | - | - |
| Day after Ascension : *Fri* | 22/05/2009 | 00:00-23:30 | - |
| *Category E: Non-bridging proximity days that precede a special day, and occur on a weekday* | | | |
| Christmas Week : *Mon* | 21/12/2009 | - | - |
| Christmas Week : *Tue* | 22/12/2009 | - | - |
| Christmas Week : *Wed* | 23/12/2009 | 05:00-23:00 | - |
| Christmas Week : *Thu* | 24/12/2009 | 00:00-23:30 | - |
| *Category F: Non-bridging proximity days that follow a special day, and occur on a weekday* | | | |
| Christmas Week : *Mon* | 28/12/2009 | 00:00-23:30 | - |
| Christmas Week : *Tue* | 29/12/2009 | - | - |
| Christmas Week : *Wed* | 30/12/2009 | - | - |
| *Category G: Non-bridging proximity days that follow a special day, and occur on a weekend* | | | |
| Christmas Week : *Sun* | 27/12/2009 | 06:30-07:30 | - |

Note: As expected, the indicator variables ($A_{h,t}$, $B_{h,t}$ and $C_{h,t}$) are non-zero for most periods on *basic special days*. This variable coding scheme reflects that load observed on basic special days is considerably lower compared to load observed on normal days and proximity days.

# 6. Benchmark Methods

We present five simple benchmarks in Section 6.1. In addition, as sophisticated benchmarks, we present Arora and Taylor's (2013) rule-based adaptations of Holt-Winters-Taylor (HWT) exponential smoothing and artificial neural networks (ANNs) in Section 6.2 and Section 6.3, respectively.

## 6.1. Simple Benchmarks

We consider five simple benchmark methods to model load for a given special day, we use: 1) **Recent Sunday** – load observed on the most recent Sunday. This benchmark method was employed by Smith (2000).

2) **Seasonal random walk (SRW)** – load observed on the same special day in the last year.

3) **Seasonal random walk for same day of the week (SRW-Day)** – load observed on the same special day in a past year, with the year chosen as the most recent year in which both current and previous special days occur on the same day of the week.

4) **Seasonal random walk for weekday/weekend (SRW-WkDay/WkEnd)** – load observed on the same special day in a past year, with the year chosen as the most recent year in which both current and previous special days belong either to a weekday, or a weekend. This benchmark method is based on the rule presented in Section 4.

5) **Seasonal random walk for same intraday cycle (SRW-IC)** – load observed on the same special day in a past year, with the year chosen as the most recent year in which both current and previous special days belong to the same intraday cycle. Following Taylor (2010), we treat a week as comprising five different intraday cycles.

## 6.2. Rule-based HWT

In this paper, we implement the rule-based HWT exponential smoothing method, proposed by Arora and Taylor (2013) for normal and anomalous load. This method focuses on modelling anomalous load in terms of its deviation from normal load. Specifically, to model French load for special days, the method employs the daily and weekly seasonal indices for normal days, and adjusts the anomalous load profile accordingly for special days using the annual seasonal index. A single source of error state space model is used, which requires the estimation of smoothing parameters for the level, intraday, intraweek and intrayear seasonal indices, along with a parameter that adjusts for first order autocorrelation in the error. The smoothing parameters determine the rate at which the level and seasonal indices are updated. Observations from the first year in the training set are used to initialize the level and seasonal indices. The value of $m_3(t)$ is selected using the same rule-based approach for special days, as used in the RB-SARMA method. The model parameters are estimated by maximum likelihood, employing the same optimization scheme and log likelihood expression used for the RB-SARMA method, whereby we use different variances for model errors for normal and special days. We refer to this method as RB-HWT.

### 6.3. Rule-based ANN

ANNs have been widely used for modelling anomalous load (see, for example, Kim *et al.*, 2000; Hippert *et al.*, 2005). In this study, we employ a feed-forward ANN method with a single hidden layer and a single output, as used by Arora and Taylor (2013). We pre-process the load data prior to modelling, using a double differencing operator of the form $(1 - L^{m_1})(1 - L^{m_2})$. We difference the load data using this operator, and normalize it by subtracting the mean, and dividing by the standard deviation. We used this variable as output, and we used lagged values of the output as inputs to the ANN. As ANNs have been shown to be unsuitable for generating multi-step ahead forecasts (Atiya *et al.*, 1999), we build a separate ANN model for each forecast horizon. We selected the lags for the input variables to be as consistent as possible with the SARMA model. Using a rule-based ANN, we model the normal and special days in a unified framework. The rule-based approach avoids the complexity of employing separate ANNs for different types of special days and normal days. The value of $m_3(t)$ is selected using the rule-based approach for special days, as used in RB-SARMA and RB-HWT. We estimate the model parameters using cross-validation, employing a hold-out sample corresponding to the last one year of the estimation sample. To estimate the link weights, we used least squares with backpropagation. We choose the activation functions for the hidden and output layer to be sigmoid and linear, respectively. We selected the input variables, number of units in the hidden layer, backpropagation learning rate and momentum parameters, and regularization parameters using only the cross-validation hold-out data. We refer to this method as RB-ANN.

## 7. Empirical Comparison

We provide an empirical comparison of the methods discussed in Sections 4, 5 and 6, based on an evaluation of their point and density forecast accuracy for the post-sample period, which consists of all half-hours in 2009. In our discussion of the results, we use the terminology, SARMA, HWT and ANN to refer to the original versions of the SARMA method, HWT exponential smoothing, and ANNs, respectively. These methods make no attempt to model the special days, and are presented by Taylor (2010). As we indicated previously, we refer to the corresponding rule-based method as RB-SARMA, RB-HWT and RB-ANN, respectively. Moreover, we refer to the SARMA method with indicator variables for special days as SARMAX(ABC). In Section 7.1, we consider point forecasting, and then discuss density forecasting in Section 7.2.

*7.1. Point Forecasting*

To evaluate point forecasts, we use the Mean Absolute Percentage Error (MAPE) and Root Mean Squared Percentage Error (RMSPE). The relative model rankings were similar for the two measures; hence, we present results using only the MAPE in this paper.

TABLE 3: MAPE ACROSS ONLY SPECIAL DAYS FOR FIVE SIMPLE BENCHMARKS, TWO SOPHISTICATED BENCHMARKS, AND THE ORIGINAL VERSION AND RULE-BASED ADAPTATION OF SARMA, AND SARMAX (ABC).

| Horizon (in hours) | 1-3 | 4-6 | 7-9 | 10-12 | 13-15 | 16-18 | 19-21 | 22-24 |
|---|---|---|---|---|---|---|---|---|
| *Simple benchmarks* | | | | | | | | |
| Recent Sunday | 11.66 | 11.66 | 11.65 | 11.65 | 11.65 | 11.65 | 11.68 | 11.71 |
| SRW | 10.49 | 10.51 | 10.53 | 10.56 | 10.58 | 10.61 | 10.63 | 10.65 |
| SRW-Day | 11.72 | 11.73 | 11.75 | 11.77 | 11.79 | 11.81 | 11.83 | 11.84 |
| SRW-WkDay/WkEnd | 10.09 | 10.11 | 10.13 | 10.15 | 10.18 | 10.20 | 10.23 | 10.25 |
| SRW-IC | 10.98 | 11.00 | 11.02 | 11.05 | 11.07 | 11.10 | 11.12 | 11.15 |
| *Sophisticated benchmarks* | | | | | | | | |
| RB-HWT | 1.35 | 2.83 | 3.88 | 4.65 | 5.23 | 5.82 | 6.38 | 6.90 |
| RB-ANN | 1.36 | 2.93 | 4.13 | 4.93 | 5.11 | 5.67 | 6.18 | 6.69 |
| *SARMA-based methods* | | | | | | | | |
| SARMA | 1.13 | 2.71 | 3.89 | 4.67 | 5.21 | 5.88 | 6.60 | 7.29 |
| RB-SARMA | **0.53** | **1.17** | **1.71** | **2.11** | **2.42** | **2.69** | **2.95** | **3.22** |
| SARMAX(ABC) | 1.12 | 2.67 | 3.82 | 4.59 | 5.14 | 5.78 | 6.44 | 7.08 |

Note: The best performing model at each horizon (i.e., best method in each column) is denoted in **bold**. Smaller MAPE values (reported in %) are better.

In Table 3, we present the MAPE across special days for the five simple benchmark methods of Section 6.1, rule-based adaptations of HWT exponential smoothing and ANNs presented in Sections 6.2 and 6.3, respectively, and SARMA-based methods of Sections 4.1 and 5. Table 3 shows that the rule-based SARMA method is considerably superior to all the other models considered in this study, at all horizons. Encouragingly, the RB-SARMA method is significantly more accurate than the SARMA method. This result justifies and highlights the importance of incorporating domain knowledge in the modelling for the French anomalous load. The forecast performances of the sophisticated benchmarks and SARMA-based univariate methods are significantly superior to the simple benchmark methods. We used the Diebold-Mariano (DM) test to verify that the difference in one-step ahead prediction accuracy between RB-SARMA and other methods was statistically significant (using a 5% significance level). We used differences of squared forecast errors to compute the test statistic.

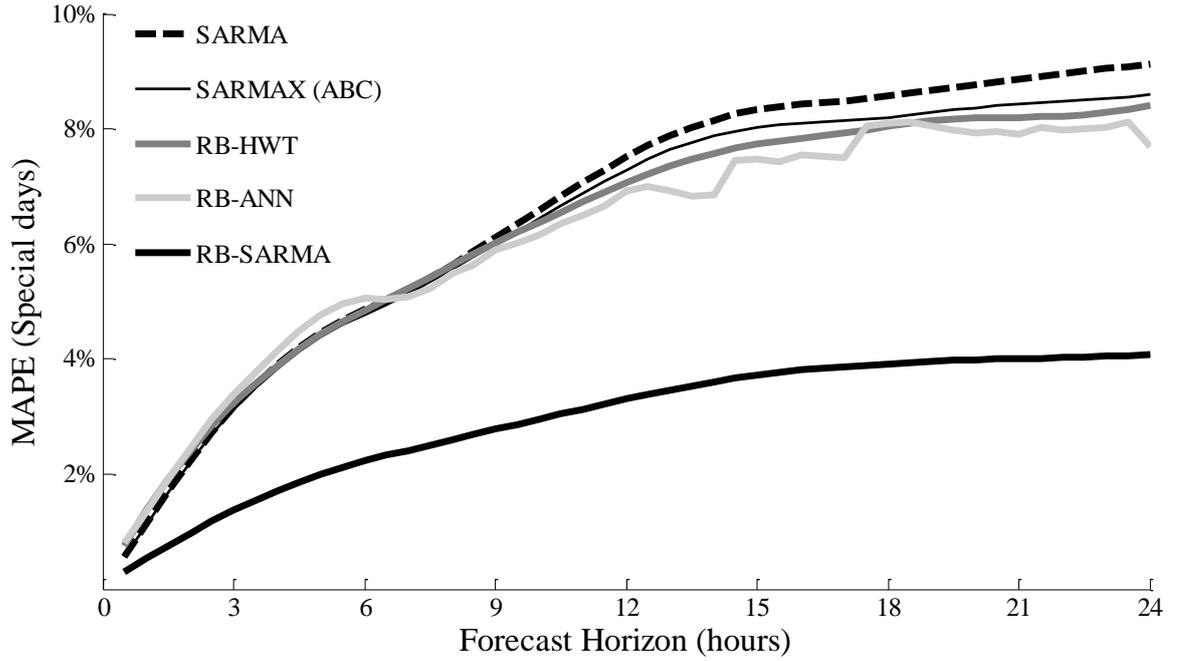

**Figure 7**— MAPE across only special days for two sophisticated benchmark methods (RB-HWT and RB-ANN), original version and rule-based adaptation of the SARMA method (SARMA and RB-SARMA), and SARMAX(ABC).

In Figure 7, we present the MAPE across special days for the best performing methods from Table 3, across all horizons. Specifically, we plot MAPEs for the SARMA-based method and the more sophisticated benchmarks. The figure emphasises what we saw in Table 3; most accurate method is RB-SARMA. It comfortably outperforms both the original SARMA method and the SARMAX(ABC) approach. The SARMAX(ABC) method was only marginally more accurate than SARMA, which makes no attempt to model the special days. Furthermore, both RB-HWT and RB-ANN performed better than SARMAX(ABC).

In Figure 8, for two chosen forecast horizons, we present the MAPE results across special days for the best performing method from Figure 7, which is RB-SARMA, and its original counterpart plotted against different times of the day. Figure 8a presents the MAPE for six-hour ahead forecasting, while Figure 8b shows the MAPE for one-day ahead prediction. As expected, the larger MAPE values correspond to periods of the day when load changes by a relatively large amount. For both forecast horizons considered in Figure 8, the accuracy of the two methods are noticeably different, especially during early morning hours (around 8 am), when the RB-SARMA method is superior.

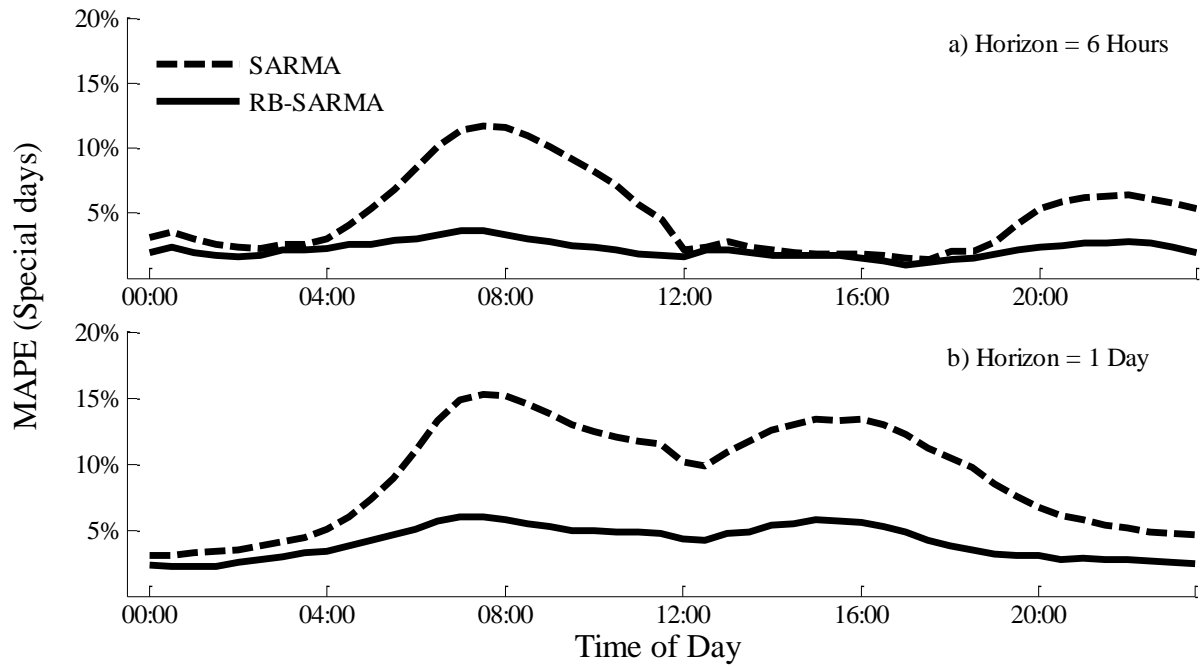

**Figure 8**— MAPE across only special days for SARMA and RB-SARMA, plotted against different times of the day, with forecast horizon equal to: a) six-hour, and b) one-day.

In Figure 9, we evaluate the methods considered in Figure 7 across all days in the one year post-sample period, i.e. across both normal and special days. We have also included in Figure 9 the original HWT and ANN methods. Interestingly, rule-based HWT performed very similarly to the original HWT method, and the same is true for rule-based ANN versus the original ANN. Figure 9 shows that RB-SARMA is considerably more accurate than all the other methods at all forecast horizons. The fact that this method outperforms the original SARMA method is impressive, and shows that careful consideration of the special days during the modelling results in an improvement in the model performance.

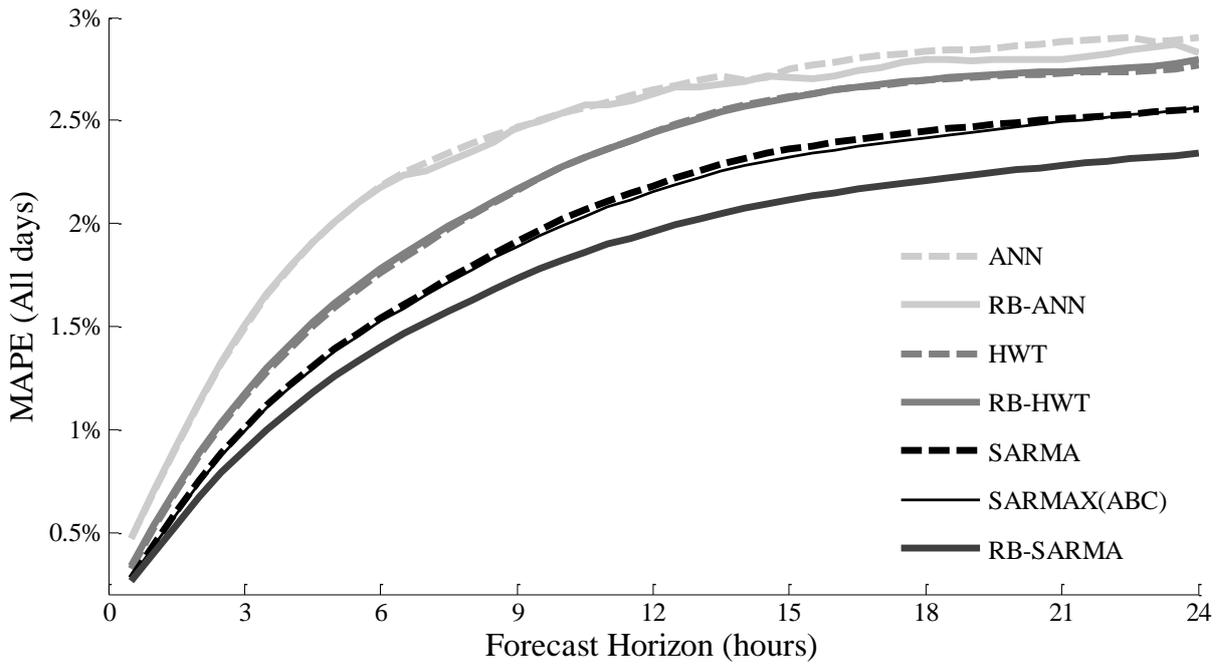

**Figure 9**— MAPE across all days for the original and rule-based adaptations of the SARMA, HWT exponential smoothing, and the ANN method, along with SARMAX(ABC).

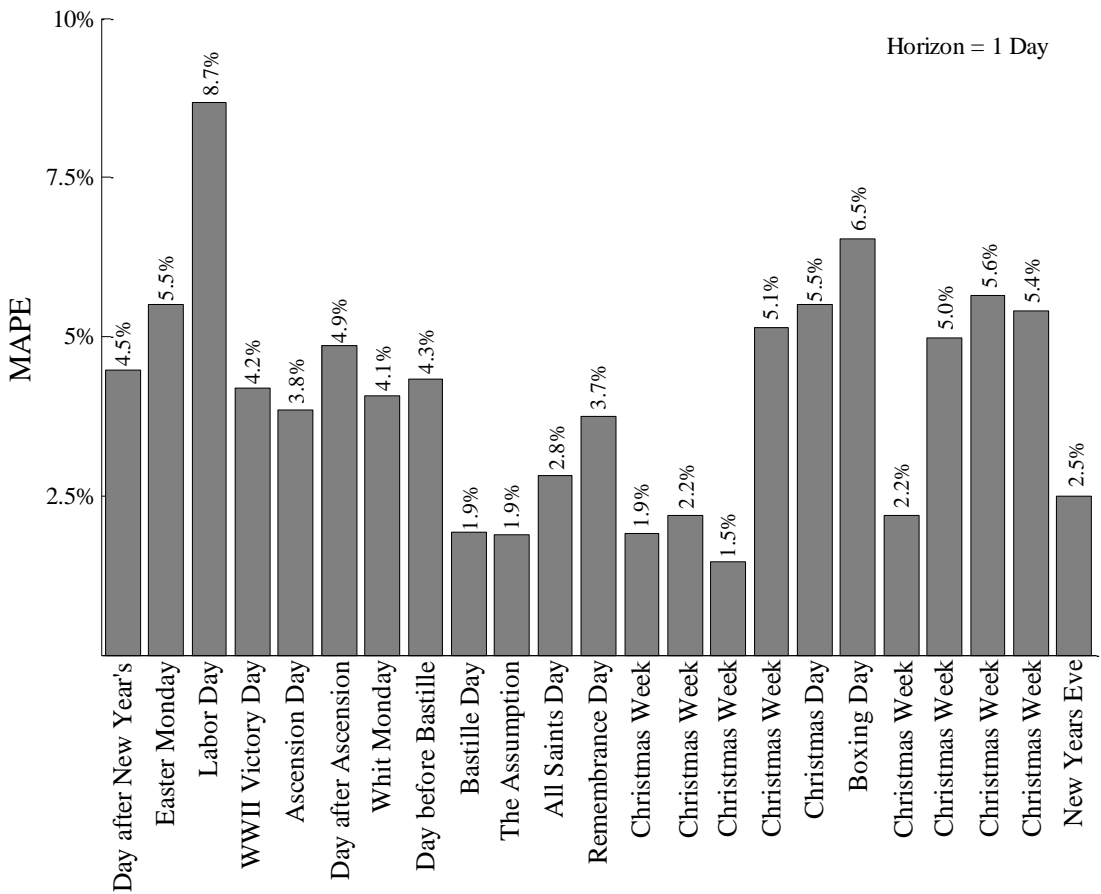

**Figure 10**— MAPE for each individual special day using the RB-SARMA method, plotted for forecast horizon corresponding to one-day (numerical MAPE values for each special day are reported above its corresponding error bar).

In Figure 10, we plot the MAPE for one-day ahead forecasting from the RB-SARMA method for the individual special days in 2009. The plot shows only 23 of the 24 special days in 2009, because we were not able to produce one-day ahead predictions for each hour on New Year's Day 2009, as we had used observations up to, and including, New Year's Eve 2008 in our estimation sample.

*7.2. Density Forecasting*

In order to evaluate the density forecast performance, we use the Continuous Ranked Probability Score (CRPS) (see, for example, Gneiting *et al*., 2007), as it takes into account both calibration and sharpness in the evaluation of the density forecasts. We generated density forecasts using SARMA and RB-SARMA using Monte Carlo simulation with 1000 iterations. Note that as opposed to SARMA, which uses one model error for all days, RB-SARMA uses different variances for model errors for normal and special days. To evaluate density forecast accuracy, in Figure 11, we present the CRPS values across special days and all days for SARMA and RB-SARMA. Note that lower CRPS values are better. It is evident from Figure 11, that RB-SARMA is considerably more accurate than SARMA, across both special days and all days, for all lead times. The model rankings based on the CRPS are consistent with the earlier rankings obtained using the MAPE.

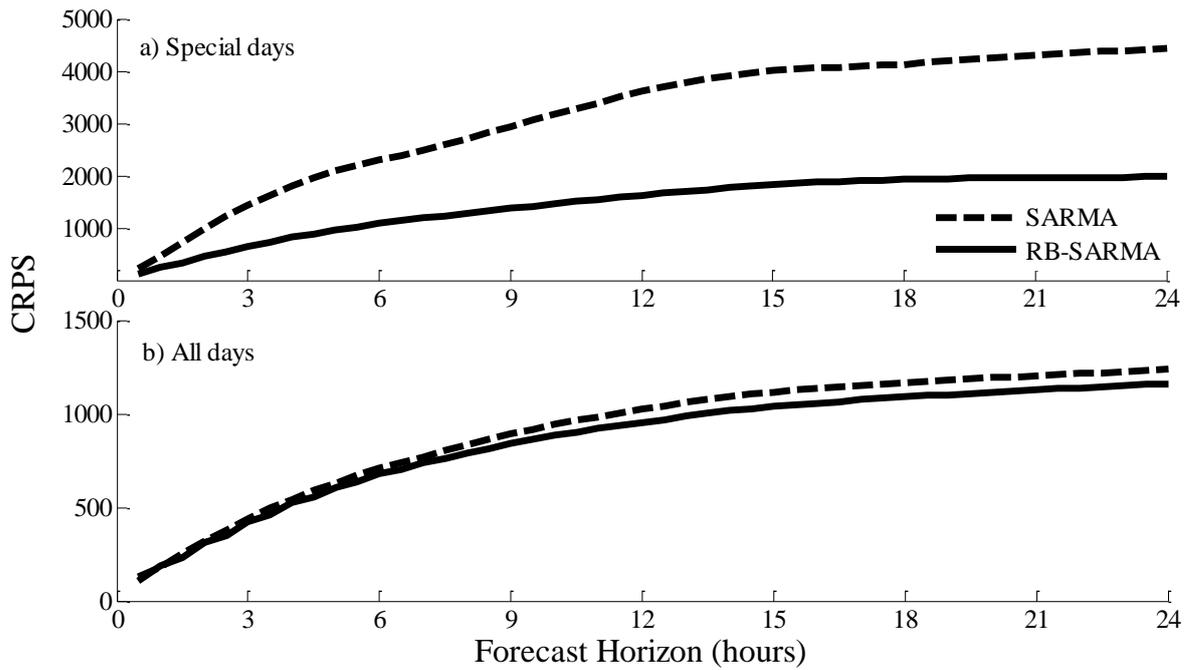

**Figure 11**— CRPS for SARMA and RB-SARMA, plotted across different horizons for only: a) Special days, and b) All days.

## 8. Summary and Concluding Remarks

In this paper, we presented a case study on load forecasting for France, with emphasis on forecasting load on special days using a rule-based SARMA method, developed by Arora and Taylor (2013) for British load data. In comparison with that study, modelling anomalous load for France is more challenging, due to the relatively large number of different types of special days in France. This extra complexity in the data necessitated our development of a new rule, formulated such that each special day is treated as having a unique profile that allows for greater flexibility during the modelling. A further methodological development in this paper is our adaptation of a SARMA method recently proposed in this journal for anomalous load (see Kim, 2013).

Overall, we found that the rule-based SARMA method generated the most accurate forecasts for special days. For these days, the MAPE obtained using rule-based SARMA was about one-third of the MAPE for the simple benchmark methods, and about a half of the MAPE of the original SARMA method, which is not rule-based and which makes no attempt to model special days. Moreover, while evaluating the probability density forecast accuracy using the CRPS, the performance of rule-based SARMA was noticeably superior to SARMA. Encouragingly, the rule-based SARMA method was considerably more accurate than rule-based HWT and rule-based ANN methods. Although the inclusion of additional indicator variable in the formulation of Kim's SARMAX method led to substantial improvements in its accuracy, it was notably outperformed by the rule-based SARMA method. One of the most encouraging findings in our study was that, in comparison with the original SARMA model, that treats special days no differently from normal days, the use of rule-based SARMA led to a noticeable improvement in accuracy when evaluated over special days.

Crucially, as opposed to some of the previous approaches, which employ different models for normal and special days, the rule-based methods investigated in this study model load for all days in a unified and coherent framework. This modelling approach makes the task of generating multistep density forecasts relatively straightforward. Moreover, the proposed methodology can potentially be adapted for other applications. Some examples where this approach could be useful includes forecasting call centre arrivals, hospital admissions, water usage, and transportation counts, as the corresponding time series exhibit seasonality and anomalous conditions, which pose significant modelling challenges. In this paper, we have considered only univariate methods. With a view to producing forecasts for longer lead times, an interesting and potentially useful line of future work would be to consider how the rule-based adaptations presented in this paper could be incorporated in a weather-based model.

## Acknowledgement

We are very grateful to the International Institute of Forecasters (IIF) and SAS for funding this research. We would also like to thank the Editor and two anonymous referees for their useful comments and suggestions.

# References

Adyaa, M., Collopy, F., Armstrong, J.S. and Kennedy, M. (2000). An application of rule-based forecasting to a situation lacking domain knowledge, *International Journal of Forecasting*, 16, 477–484.

Armstrong, J.S. (2001). Principles of forecasting: A handbook for researchers and practitioners. Kluwer Academic Publishers, Boston.

Armstrong, J.S. (2006). Findings from evidence-based forecasting: Methods for reducing forecast error, *International Journal of Forecasting*, 22, 583–598.

Arora, S. and Taylor, J.W. (2013). Short-term forecasting of anomalous load using rule-based triple seasonal methods, *IEEE Transactions on Power Systems*, 28, 3235–3242.

Arora, S. and Taylor, J.W. (2016). Forecasting electricity smart meter data using conditional kernel density estimation, *Omega*, 59, 47–59.

Atiya, A.F., El-Shoura, S.M., Shaheen, S.I. and El-Sherif, M.S. (1999). A comparison between neural-network forecasting techniques–case study: river flow forecasting, *IEEE Transactions on Neural Networks*, 10, 402–409.

Barrow, D. and Kourentzes, N. (2016). The impact of special days in call arrivals forecasting: A neural network approach to modelling special days, *European Journal of Operational Research*, in press.

Box, G.E.P. and Jenkins, G.M. (1970). Time series analysis, forecasting and control, San Francisco: Holden–Day.

Bunn, D.W. (1982). Short-term forecasting: A review of procedures in the electricity supply industry, *The Journal of the Operational Research Society*, 33, 533–545.

Bunn, D.W. (2000). Forecasting loads and prices in competitive power markets, *Proceedings of the IEEE*, 88, 163–169.

Cancelo, J.R., Espasa, A. and Grafe, R. (2008). Forecasting the electricity load from one day to one week ahead for the Spanish system operator, *International Journal of Forecasting*, 24, 588–602.

Cho, H., Goude, Y., Brossat, X. and Yao, Q. (2013). Modeling and forecasting daily electricity load curves: a hybrid approach, *Journal of the American Statistical Association*, 108, 7–21.

Collopy, F. and Armstrong, J.S. (1992). Rule-based forecasting: development and validation of expert systems approach to combining time series extrapolations, *Management Science*, 38, 1394–1414.

Cottet, R. and Smith., M. (2003). Bayesian modelling and forecasting of intraday electricity load, *Journal of the American Statistical Association*, 98, 839–849.

De Livera, A.M., Hyndman, R.J. and Snyder, R.D. (2011). Forecasting time series with complex seasonal patterns using exponential smoothing, *Journal of the American Statistical Association*, 106, 1513–1527.

Dordonnat, V., Koopman, S.J., Ooms, M., Dessertaine, A. and Collet, J. (2008). An hourly periodic state space model for modelling French national electricity load, *International Journal of Forecasting*, 24, 566–587.

Gneiting, T., Balabdaoui, F. and Raftery, A.E. (2007). Probabilistic forecasts, calibration and sharpness, *Journal of the Royal Statistical Society*, 69, 243–268.

Gould, P.G., Koehler, A.B., Ord, J.K., Snyder, R.D., Hyndman, R.J. and Vahid-Araghi, F. (2008). Forecasting time series with multiple seasonal patterns, *European Journal of Operational Research*, 191, 207–222.

Hippert, H.S., Bunn, D.W. and Souza, R.C. (2005). Large scale neural networks for electricity load forecasting: Are they overfitted?, *International Journal of Forecasting*, 21, 425–434.

Hyde, O. and Hodnett, P.F. (1993). Rule-based procedures in short-term electricity load forecasting, *IMA Journal of Mathematics Applied in Business and Industry*, 5, 131–141.

Hyde, O. and Hodnett, P.F. (1997). An adaptable automated procedure for short-term electricity load forecasting, *IEEE Transactions on Power Systems*, 12, 84–94.

Kim, K-H., Youn, H-S. and Kang, Y-C. (2000). Short-term load forecasting for special days in anomalous load conditions using neural networks and fuzzy inference method, *IEEE Transactions on Power Systems*, 15, 559–565.

Kim, M.S. (2013). Modeling special-day effects for forecasting intraday electricity demand, *European Journal of Operational Research*, 230, 170–180.

Lagarias, J. C., Reeds, J. A., Wright, M. H. and Wright, P. E (1998). Convergence properties of the Nelder-Mead simplex method in low dimensions. *SIAM Journal of Optimization*, 9, 112–147.

Pardo, A., Meneu, V. and Valor, E. (2002). Temperature and seasonality influences on Spanish electricity load, *Energy Economics*, 24, 55–70.

Rahman, S. and Bhatnagar, R. (1988). An expert system based algorithm for short term load forecast, *IEEE Transactions on Power Systems*, 3, 392–399.

Ramanathan, R., Engle, R., Granger, C.W.J., Vahid-Araghi, F. and Brace, C. (1997). Short-run forecasts of electricity load and peaks, *International Journal of Forecasting*, 13, 161–174.

Smith, M. (2000). Modeling and short-term forecasting of New South Wales electricity system load, *Journal of Business Economics and Statistics*, 18, 465–478.

Soares, L.J. and Medeiros, M.C. (2008). Modelling and forecasting short-term electricity load: A comparison of methods with an application to Brazilian data, *International Journal of Forecasting*, 24, 630–644.

Taylor, J.W. (2008). An evaluation of methods for very short-term load forecasting using minute-by-minute British data, *International Journal of Forecasting*, 24, 645–658.

Taylor, J.W. (2010). Triple seasonal methods for electricity demand forecasting, *European Journal of Operations Research*, 204, 139–152.

Weron, R. (2006). *Modelling and Forecasting Electric Loads and Prices*. Chichester, Wiley.